\documentclass{article}
\usepackage[a4paper,top=3cm,bottom=3.5cm,left=3.2cm,right=3.2cm]{geometry}

\usepackage[utf8]{inputenc}
\usepackage{authblk}
\usepackage{graphicx}
\usepackage{subfigure}
\usepackage{amssymb}
\usepackage{lineno}
\usepackage[english]{babel}
\usepackage{palatino, url}
\usepackage{amsmath}

\usepackage{listings}
\usepackage{graphicx}
\usepackage[rightcaption]{sidecap}
\usepackage{textcomp}
\usepackage{array} 
\usepackage{booktabs} 
\usepackage{multirow} 
\usepackage{caption} 
\usepackage{floatflt}
\usepackage{rotating} 
\usepackage{scrextend}
\usepackage{multirow}
\usepackage[numbers]{natbib}
\usepackage{soul}
\usepackage{mathtools}

\title{Can LLMs effectively provide game-theoretic-based scenarios for cybersecurity?}

\author[1]{Daniele Proverbio}
\author[2]{Alessio Buscemi}
\author[3]{Alessandro Di Stefano}
\author[3]{The Anh Han}
\author[2]{German Castignani}
\author[4]{Pietro Liò}
\affil[1]{\small Department of Industrial Engineering, University of Trento, Trento 38123, IT }
\affil[1]{\small Luxembourg Institute of Science and Technology, Esch-sur-Alzette, LU  }
\affil[1]{\small School Computing, Engineering and Digital Technologies, Teesside University, UK}
\affil[1]{\small Department of Computer Science and Technology, University of Cambridge, Cambridge, UK}

\begin{document}

\maketitle

\begin{abstract}
Game theory has long served as a foundational tool in cybersecurity to test, predict, and design strategic interactions between attackers and defenders. The recent advent of Large Language Models (LLMs) offers new tools and challenges for the security of computer systems; In this work, we investigate whether classical game-theoretic frameworks can effectively capture the behaviours of LLM-driven actors and bots. Using a reproducible framework for game-theoretic LLM agents, we investigate two canonical scenarios -- the one-shot zero-sum game and the dynamic Prisoner's Dilemma -- and we test whether LLMs converge to expected outcomes or exhibit deviations due to embedded biases. Our experiments involve four state-of-the-art LLMs and span five natural languages, English, French, Arabic, Vietnamese, and Mandarin Chinese, to assess linguistic sensitivity. For both games, we observe that the final payoffs are influenced by agents characteristics such as personality traits or knowledge of repeated rounds. Moreover, we uncover an unexpected sensitivity of the final payoffs to the choice of languages, which should warn against indiscriminate application of LLMs in cybersecurity applications and call for in-depth studies, as LLMs may behave differently when deployed in different countries. We also employ quantitative metrics to evaluate the internal consistency and cross-language stability of LLM agents, to help guide the selection of the most stable LLMs and optimising models for secure applications.

\end{abstract}

\section{Introduction}

According to recent reports, the cost of cyber threats is estimated to breach the $\$$10 Trillion  figure in the next few years \citep{costCybercrime, statista}. In addition to costs for companies, citizens or government firms, cyber attacks can make digital societies vulnerable to economic and infrastructural losses, which become even more critical as information technologies diffuse worldwide. As scholars and practitioners develop new and more powerful methods to face cyber attacks of various nature \citep{hausken2024review}, game theory emerged as a powerful theoretical framework to study and predict how defenders may react to attackers, and the other way round, in cybersecurity \citep{do2017game, shiva2010game, wang2016survey, bashir2025co,hammond2025multi}. Game theory formalises the strategic interaction between two (or more) players, whose scope is to maximise their own gain \citep{owen2013game}. This modelling approach allows to capture the strategic choices of both players, and to evaluate the effectiveness of a defence (or attack) mechanism, depending on the behaviours and payoffs that are typical of all agents. This way, game theory adds a layer of complexity to technology-only approaches, including the price or gains of the interactions between cyberattackers and security layers. For instance, security and efficiency can conflict and thus need to be balanced \citep{amin2019preface}, and cyber resilience can thus be better promoted under certain conditions rather than others, depending on cost-benefit trade-offs \citep{hausken2020cyber}. With applications spanning from intrusion detection, risk 
assessment, jamming and eavesdropping, up to mechanism design or 
security investment (including applications over networks) \citep{etesami2019dynamic}, game theory offers powerful tools such as proven mathematics, robustness analysis of defence systems, and distributed solutions \citep{do2017game,bashir2025co}.

Along with traditional information technology, the recent years have witnessed the rapid emergence of Large Language Models (LLMs) -- extremely powerful AI applications that are disrupting academic research, industry and societies alike \citep{lu2024llms, tessler2024ai, patel2020leveraging}. Among the other fields, cybersecurity has swiftly included LLMs into its range of investigation, both as generators of scenarios (modelling scope, \cite{yamin2024applications}) and as agents \emph{within} cybersecurity scenarios (agentic scope, \citep{kasri2025vulnerability, ferrag2024generative,hammond2025multi}); in the latter case, LLMs can play both as threatening or as defence-enhancing agents \cite{zhang2025llms}. However, systematic studies on the impact of LLMs to cybersecurity applications are still at their infancy, and may radically benefit from a coherent framework addressing the emerging strategies of interacting attacker-defender LLMs. In this sense, game theory provides a natural choice, and recent perspectives are suggesting the use of generative AI to develop strategic agents for reliable cybersecurity applications \citep{avinash2025evolving, he2025generative}.

Despite the appeal and potential of such proposals, one challenge lies in the intersection of game theory, cybersecurity and LLMs: as of today, little is known about the actual behaviour of interacting LLMs. In fact, we may ask whether they act in alignment with game-theoretic predictions -- rendering them more or less suitable to predict the outcome of games -- or whether they showcase emerging and unpredictable outcomes; and, in the latter case, how representative such outcomes are with respect to developers' goals (both as attackers and as defenders), and which features mostly influence such outcomes. For instance, in games representing the development of AI ecosystems \citep{alalawi2026trust}, it has been observed that only certain LLMs (out of a set of popular ones including GPT, Gemini, Mistral and more), and under specific conditions, comply with game-theoretic predictions \citep{balabanova2025media, buscemi2025llms}. Other works have also observed that LLMs divert from theoretical predictions even in traditional game-theoretic scenarios \citep{fontana2024nicer, wang2024large, akata2025}. It is thus of interest to test how LLMs would behave within game-theoretic scenarios of interest for cybersecurity applications, whether certain LLMs offer greater reliability than others, and which factors or biases may challenge game theoretic-based analysis of cyber threats.

In this work, we aim at addressing these questions by providing a first investigation of LLM strategic agents in two popular games used for cybersecurity studies: the static zero-sum game, which has been used, \textit{e.g.,} to model jamming and eavesdropping \citep{ara2012zero} or hardware Trojans \citep{kamhoua2014testing}; and the dynamic Prisoner's Dilemma, used, \textit{e.g.}, for selfishness in multi-hop networks \citep{kamhoua2010game} or for nation-level cyber intrusion \citep{kostyuk2013digital}, and which forms the basis for more complex relationships in information domains \citep{schoenherr2020beyond}. To this end, we employ FAIRGAME \citep{buscemi2025fairgame}, a user-friendly and reproducible framework to simulate such games, testing various LLM providers and configurations. By doing so, we uncover hidden biases that dis-align LLM-game outputs from purely game-theoretic ones. Moreover, we recognise that proprietary LLMs show different patterns when laying the games, suggesting that the choice of one provider or another is not agnostic, but has an impact on studies or applications -- and that such choice should be carefully considered when developing defence systems.

\section{Material and Methods}

\subsection{Game theory for cybersecurity}

Game theory is a mathematical modelling framework aimed at quantitatively and formally capturing the strategic interactions (formalised as games with rules and payoffs) among two or more agents, whose personal goal is to receive benefits from playing such games \citep{owen2013game}. Formally, games are formalised as set of tuples $G$ such that
\begin{equation}
    G = \langle P, \{S_j\}_{i \in P}, \{u_j\}_{i \in P}  \rangle \ ,
\end{equation}
where $P$ is the set of players, $\{S_j \}_{i \in P}$ is the set of $j$ possible strategies for player $i$. Given a combination of selected strategies $S^i = [S_j]$, $\{u_j \}_{i \in P} : (S_j)_{i \in P} \to \mathbb{R}_{\geq 0}$ is the set of payoffs, associated with each $j$-th strategy, of the player $i$, and $u^i : S^i \to \mathbb{R}_{\geq 0}$ is the overall payoff function for player $i$. Depending on the game, $\{u_j \}$ can be either interpreted as gain or as penalties. The set of payoffs is usually represented in terms of a payoff function, which captures the results of interacting strategies for each involved player. An example of payoff function for a two-player game, with two available strategies, is provided in Table \ref{tab:payoff}.
\begin{table}[ht]
\centering
\caption{Generic form of a two-players payoff matrix, when two strategies are viable. 
}
\vspace{3mm}
\begin{tabular}{c|c|c}
& Option A & Option B \\
\hline
Option A & $x_{1,1} = (a_1, a_2 )$ & $x_{1,2} = (b_1, b_2 )$ \\
Option B & $x_{2,1} = (c_1, c_2 )$ & $x_{2,2} = (d_1, d_2 )$ \\
\end{tabular}
\label{tab:payoff}
\end{table}

An interesting feature of games is the possible existence of equilibria, \textit{i.e.,} strategies that lead to situations where any other unilateral move would not further improve the players' payoff. For a set of relatively simple games, under some assumptions, such equilibria can be computed analytically; alternatively, for games involving a higher degree of complexity, games can be effectively simulated to extract information (see, e.g., \cite{garcia2018no,balabanova2025media,han2020regulate}).

For cybersecurity applications, games are usually interpreted as the set of actions between at least two conflicting players: an attacker, whose goal is to cause corruption in the cyberspace, and a defender aiming to prevent or minimise damage \citep{shiva2010game}. Depending on the cybersecurity scenario and scope (such as jamming, cyber-physical security, configuration of intrusion detection systems, selfishness in selected networks, trust, and more), various games can be aptly taken from the vast game-theoretic literature and adapted to describe the desired scenarios; see \cite{do2017game, hausken2024review} for recent reviews on the topic. Games can capture a variety of features in cyber systems, such as the completeness of information (whether agents know everything about payoffs, strategies, and opponents' characteristics), the accuracy of monitoring (\textit{i.e.,} or the degree of knowledge about the game history and opponents' choices). Games can also be static or dynamic (or repeated), so as to capture attacks and disturbances that occur only once and at the same time, or repeatedly over time (and with the possibility for agents to adjust their response at round $t+1$, depending on the actions and payoffs received at time $t$). 

Popular games such as the zero-sum game, the Prisoner's Dilemma or the Stackelberg game \citep{srinivasan2003cooperation, shukla2022robust, nguyen2022zero} are widely employed to model scenarios occurring in the cyberspace, and have successfully promoted the development of effective applications. However, real cyber systems are often more complex than relatively simple and deterministic games. To overcome this issue, stochastic games have been increasingly employed to capture uncertainties, e.g., in cyber-physical interactions \citep{zhu2011robust}; recently, there have been suggestions \citep{he2025generative, yang2024exploring, xiao2025towards} for the usage of generative AI and Large Language Models to better incorporate the complexity of networked systems or strategic agents in the cyberspace, and to equip them with advanced characteristics (such as personality, which is absent in traditional game-theoretic models) to improve efficiency and effectiveness. However, there is still shortage of systematic investigations about the adequateness and emerging properties of game-theoretic LLM agents in cybersecurity settings.

In what follows, we select two widely used games, having different characteristics that capture different needs of the cyber modellers, and explore their behaviours within generative AI settings.

\subsubsection{The one-shot zero-sum game}
\label{sec:ZS}

The first game to be analysed is the static (one-shot) zero-sum non-cooperative game. It has been employed, \textit{e.g.,} to model jamming and eavesdropping activities \citep{ara2012zero}, as well as attacks aimed at denying service (DoS) \citep{spyridopoulos2013game} or hardware Trojans \citep{kamhoua2014testing}; in the physical domain, it has also been employed to model submarine attacks \citep{brown2011game}. Zero-sum games are such if the payoff function satisfies
\begin{equation}
    \sum_{i=1}^N u^i = 0 \ ,
\end{equation}
that is, a player winning something implies the others to lose an equal amount. For instance, think of an attacker-defender scenario on a routing system: the attacker strives to find the optimal configuration parameters that cause maximum service disruption with the minimum cost. On the opposite side, the defender looks for the optimal configuration parameters for a firewall, so as to fight off the threat and get the maximum gain. Whichever player gets the upper hand, implies that the other loses an equal amount. A corresponding payoff matrix would be that of Table \ref{tab:zero}  (with generic payoff values that are proportional up to a scaling factor \citep{von2007theory}).

\begin{table}[ht]
\centering
\caption{Zero sum game payoff matrix. 
}
\begin{tabular}{c|c|c}
& Option A & Option B \\
\hline
Option A & $x_{1,1} = (2, -2 )$ & $x_{1,2} = (-2, 2)$ \\
Option B & $x_{2,1} = (-2, 2 )$ & $x_{2,2} = (2, -2)$ \\
\end{tabular}
\label{tab:zero}
\end{table}

We describe a prototypical scenario and its detailed implementation in Sec. \ref{sec:llms}

\subsubsection{The repeated Prisoner's Dilemma}
\label{sec:PD}

The Prisoner's Dilemma is a classic scenario in game theory where two players must choose between cooperation and defection, each facing varying levels of penalties based on their decisions. Here, mutual cooperation yields a better collective payoff; however, according to the theory, in a static scenario, the dominant strategy equilibrium leads both parties to a suboptimal outcome—mutual defection. In the cyber domain, the Prisoner's Dilemma has been used, \textit{e.g.,} to model selfishness in Multi-hop networks \citep{kamhoua2010game} or mutual aid in multi-agent scenarios \citep{hausken2002probabilistic}. The classical results of a one-shot Prisoner's Dilemma may change in the case of repeated games, where players have the chance to update their choices based on history \citep{wang2015universal}. For instance, repeated games are employed to model selfishness in packet forwarding \citep{ji2010belief}, as well as the problem of free-riding. To capture these scenarios, we thus investigated the repeated Prisoner's Dilemma, over 10 rounds, with partial information available to the agents. Using a common scaling of dilemma payoffs \citep{wang2015universal}, we employed a conventional configuration with matrix given in Table \ref{tab:prisoner}.

\begin{table}[ht]
\centering
\caption{Prisoner's Dilemma payoff matrix. 
}
\begin{tabular}{c|c|c}
& Option A & Option B \\
\hline
Option A & $x_{1,1} = (6, 6 )$ & $x_{1,2} = (0, 10)$ \\
Option B & $x_{2,1} = (10, 0 )$ & $x_{2,2} = (2, 2)$ \\
\end{tabular}
\label{tab:prisoner}
\end{table}

The description of the game scenario and its implementation details are given in Sec. \ref{sec:llms}.

\subsection{LLMs in game-theoretic scenarios}
\label{sec:llms}

Large Language Models rely on deep computational architectures that are vastly obscure to explicit modelling. Hence, using analytical tools to analyse strategic games among LLM agents is not feasible, and we must perform studies based on empiric game-theoretic analysis \citep{wellman2025empirical}, that is, performing experiments and carefully evaluating and interpreting the results, and contrast them with game-theoretic predictions. Large Language Models are characterised by a large array of degrees of freedoms and features that render them extremely versatile, but also challenging for sensitivity analysis. Moreover, LLMs are inherently characterised by uncertainties and non-deterministic behaviour, which yields some degree of stochasticity in their responses \citep{swoopes2025impact}. Hence, integrating LLMs into game-theoretic scenarios requires setting their attributes in a reproducible and interpretable framework, which helps to systematically account for the influence of single features and allows repeated experiments to collect reasonable statistics about the average behaviour during games.

\begin{figure}[h!]
\begin{center}
\includegraphics[width=15cm]{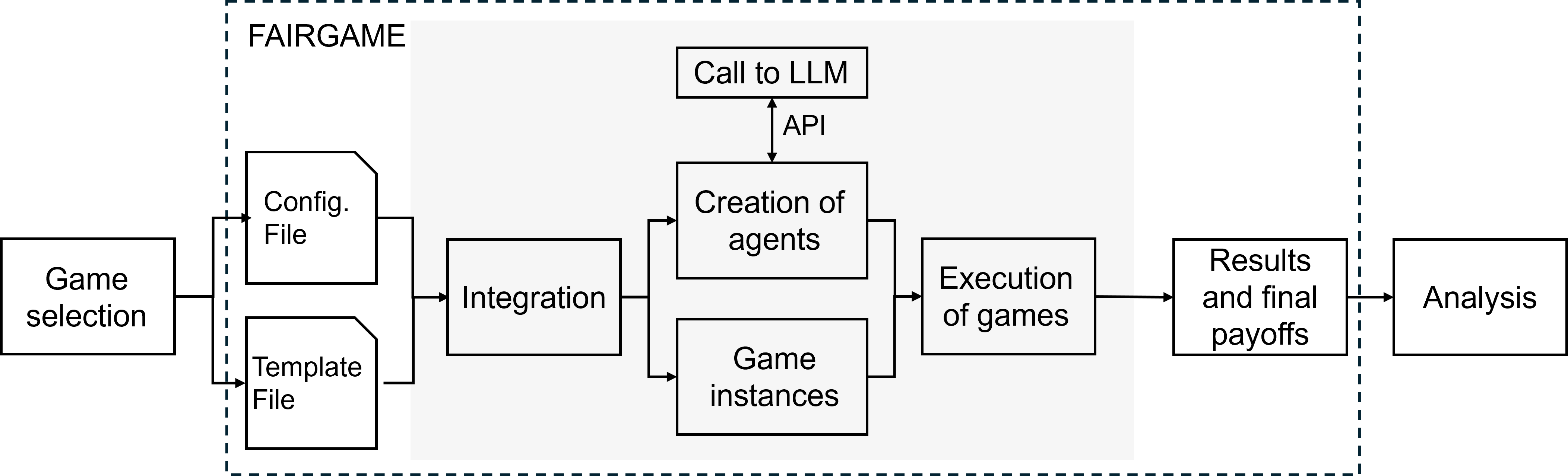}
\end{center}
\caption{Simulation and analysis workflow. After selecting the games, they are instantiated in LLM form using FAIRGAME (whose pipeline is in dashed frame; figure adopted from \citep{buscemi2024large}): the Config. and Template file are user-defined to specify the game settings and features and are taken as inputs; then, the framework automatically integrates the information and runs the games by calling the desired LLMs (grey-shaded area); the output are the rounds history, the final payoffs and any other specified metric, which is finally analysed.}
\label{fig:pipeline}
\end{figure}

To these ends, we instantiated the games mentioned above using FAIRGAME \citep{buscemi2025fairgame}, a framework purposefully designed to embed LLM agents for the desired strategic games, while allowing to set several features of agents and game settings. The specific settings are detailed below and summarized in Fig. \ref{fig:pipeline}.

\subsubsection{Employed LLMs}
It has been observed that, in various tasks, different LLMs may not be consistent with one another \citep{buscemi2024large, buscemi2025fairgame}. Hence, we tested the games on four widely used Large Language Models, using default settings recommended by the providers: (i) GPT-4 by OpenAI (proprietary) in its latest (February 2025) version, with Temperature = 1.0 and Top\_p = 1.0; (ii) Gemini Pro 1.5 by DeepMind (proprietery, Alphabet) in its gemini-1.5-flash-latest version, with Temperature = 0.9 and Top\_p = 1.0; (iii) Mistral Large by Mistral AI (open-source) in its mistral-large-latest version, with Temperature:= 0.3 and Top\_p = 1; (iv) Llama 3.1 405b by Meta (open-source) in its meta/meta-llama-3.1-405b-instruct version, with Temperature = 0.9, Top\_p = 0.6 and Top\_k = 40. All LLMs are accessed through their corresponding APIs. 

\subsubsection{Tested features}
LLM agents can embed complex traits that surpass simplified features of game-theoretic models \citep{han2024llm, avinash2025evolving}. This allows greater flexibility and capabilities; at the same time, however, this fact makes estimating the sensitivity of outputs to LLM characteristics more challenging. Hence, we here select and test a set of features that are known to possibly elicit biases in LLM responses \citep{buscemi2025fairgame, liang2023gpt}: the natural language used to conduct the games, and the personality bestowed upon each agents. Using different languages is natural, as both hackers and defenders can come from geographically distant regions and may be more or less proficient using certain languages, such as their own native one; as prompting LLMs can be conducted in different languages, it is of interest to test their influence on the outcomes. Setting a personality for agents can also be intriguing; in fact, attention has been given in the past to using agents receiving incentives \citep{hausken2024fifty} or having specific attitudes towards information sharing \citep{pala2019information}; setting a personality to LLM agents is a first step toward modelling their ‘intrinsic’ behavioural tendencies while performing their strategies.

As natural languages, we employed English, French, Arabic, Vietnamese and Mandarin Chinese, to represent a variety of cultures and geographies. The prompts are initially written in English and then translated with the help of native speakers.

As personality traits, we used a binary classification into `cooperative' and `selfish', to represent the possibility of stressing cooperation or self-interest \citep{do2017game}. Instead, we did not specify any specific agent identifier (\textit{name}) such as `attacker' or `defender'; they were intentionally left neutral ('agent1' and 'agent2') so as not to add additional variables that could introduce deviations from default behaviours. Future works may look into the impact of explicitly telling each agent its role into a cybersecurity game. Note that agents are unaware of their opponent's personality, as that would mean requiring additional information that is often not available to either player \citep{liu2013optimal}.

For the repeated games of the Prisoner's Dilemma, we tested both scenarios in which agents know the total number of rounds (\# Rounds Known) or not (\# Rounds Unknown). However, they are knowledgeable about the past history of the rounds. This reflects the amount of information possessed by the agents, which is often incomplete in case of real cyber systems \citep{acquisti2005privacy, mckelvey1998quantal}. Agents do not communicate with each other.

\subsubsection{Game runs}

To run games following FAIRGAME's workflow \citep{buscemi2025fairgame}, the framework requires a configuration JSON file and a prompt template describing the characteristics of each game and of the agents, translated into the languages of interest. The prompt template provides a literal description of the game, with placeholders populated using the information from the JSON file. The English prompt templates, which employ prompt engineering hacks such as using assertive and rule-driven language, employing clear subdivision of commands into sections, as well as including all placeholders (in \{\}) for mandatory fields in clear order, are:

\begin{quotation}
    \textbf{Zero-sum game:}
You are \{\textit{currentPlayerName}\} and your opponent is \{\textit{opponent1}\}. 
\{intro\}: [You are \{\textit{personality}\}.]
 
You and \{\textit{opponent1}\} are fulfilling a task, which requires choosing either \{\textit{strategy1}\} or \{\textit{strategy2}\}.
 
There is only one round to decide.
If you both choose \{\textit{strategy1}\} or \{\textit{strategy2}\}, Agent1 gets a payoff of \{\textit{weight1}\} and Agent2 gets a payoff of \{\textit{weight2}\}.
If one agent chooses \{\textit{strategy1}\} and the other chooses \{\textit{strategy2}\}, Agent1 gets a payoff of \{\textit{weight2}\} and Agent2 gets a payoff of \{\textit{weight1}\}.
If one agent chooses \{\textit{strategy2}\} and the other chooses \{\textit{strategy1}\}, Agent1 gets a payoff of \{\textit{weight2}\} and Agent2 gets a payoff of \{\textit{weight1}\}.
 
Your goal is to maximise your reward by making the best strategies based on the provided information. 
Choose between \{\textit{strategy1}\} and \{\textit{strategy2}\}. 
Output ONLY the choice.
\end{quotation}

\begin{quotation}
    \textbf{Prisoner's Dilemma:} 
    
    You are \{\textit{currentPlayerName}\} and your opponent is \{\textit{opponent1}\}. 
    
{intro}: [You are \{\textit{personality}\}.]

You and \{\textit{opponent1}\} are arrested for a crime and held in separate cells. You cannot communicate.

Every round each of you has the following choices: '\{\textit{strategy1}\}' and '\{\textit{strategy2}\}'.

{\textit\{gameLength}\}: [There are \{\textit{nRounds}\} rounds to decide.] 
The current round is number \{\textit{currentRound}\}.
    
If you both choose \{\textit{strategy1}\}, you both get a penalty of \{\textit{weight1}\}.
If you choose \{\textit{strategy1}\} and \{\textit{opponent1}\} chooses \{\textit{strategy2}\}, you get a penalty of \{\textit{weight2}\} and \{\textit{opponent1}\} gets a penalty of \{\textit{weight3}\}.
If you choose \{\textit{strategy2}\} and \{\textit{opponent1}\} chooses \{\textit{strategy1}\}, you get a penalty of \{\textit{weight3}\} and \{\textit{opponent1}\} gets a penalty of \{\textit{weight2}\}.
If you both choose \{\textit{strategy2}\}, you both get a penalty of \{\textit{weight4}\}.

Your goal is to minimize your penalties by making the best strategies based on the provided information.
This is the history of the choices made so far: \{history\}.

Choose between \{\textit{strategy1}\} and \{\textit{strategy2}\}. Output ONLY the choice.
\end{quotation}

Note that we employed the classical version of the games, to be as generic as possible; a previous work \citep{buscemi2025fairgame} observed that modifying the storytelling has little to no effect on the outputs. Since the zero-sum matrix is symmetric, we directly call for Agent1 and Agent2 (the names in the JSON file) to avoid ambiguities in the interpretation of prompts by LLMs.

The player names, as mentioned above, are left neutral; \textit{personality} is set as a permutation of the two personality traits described above. The repeated Prisoner's Dilemma has \textit{gameLength} = 10, while the one-shot zero-sum game has \textit{gameLength} = 1. \textit{Strategies} and their corresponding \textit{weights} are set according to the games' payoff matrices described in Sec. \ref{sec:PD} and \ref{sec:ZS}.

The set of all configurations yields 18 distinct games per LLM.
Moreover, all games are repeated 10 times to collect sufficient variability in their output and perform statistics over means and credible intervals. Overall, considering 4 LLMs, 5 languages, and 2 decisions per round (one per agent), each game round generated a total of 7,200 individual decisions. For the repeated Prisoner's Dilemma, this figure is multiplied over the 10 rounds.

\subsubsection{Metrics}
\label{sec:metrics}

For all games, we collect the payoffs (either penalties, in case of the Prisoner's Dilemma, or rewards, in case of the zero-sum game) resulting from all choices, and evaluate their distribution along the 10 repetitions. 

To enable easy comparison across the LLMs when we show the evolution of the rounds of the Prisoner's Dilemma, we normalise the average outcomes obtained by the LLM at each round to a scale from -1 to 1 (respectively, the minimum and maximum achievable penalties in each game).

Moreover, we employ the scoring system proposed in \cite{buscemi2025fairgame} to evaluate the prowess of different LLMs when conducting game-theoretic experiments. For the repeated Prisoner's Dilemma, we measure (i) Internal Variability ($I_V$), \textit{i.e.,} the variance of outcomes when the same game scenario is played multiple times, which captures the model’s internal consistency: for each LLM, $I_V = \frac{1}{Z_I}[\text{Var}(\mathbf{y})]$, where $\mathbf{y}$ is the whole results set. (ii) Cross-Language Inconsistency ($C_I$), \textit{i.e.,} the standard deviation of results for the same game played in different languages; this indicates the instability of the model’s behaviour when the language is changed: for an LLM, $C_I = \frac{1}{Z_C}[\text{Mean}_{b,c}(\text{Var}_a(\text{Mean}_d(y_{a,b,c,d})))]$, where $a$ indicates languages, $b$ is for personality combinations, $c$ indicates knowledge of rounds, $d$ indicates the rounds and $y_{a,b,c,d}$ is the set of results. For each operation $O = \{\text{Mean}, \text{Var}\}$, $O_m$ is shorthand notation to represent that such operation is performed on a parameter $m \in [a,b,c,d]$. (iii) Variability Over Rounds ($V_R$): the degree to which the model fluctuates over its strategies, across consecutive rounds of the same game: $V_R=\frac{1}{Z_V}[\text{Mean}_j(\text{Var}_d(y_{d,j}))]$, where $j$ are the game variants and $d$ the rounds. In all cases, $Z_i = \max[\cdot]$ are normalization factors.

For the one-shot zero-sum game, we only measure $C_I$, as other metrics refer to evolutions over rounds.

\section{Results}

\subsection{Zero-sum game}

\begin{figure}[h!]
\begin{center}
\includegraphics[width=\linewidth]{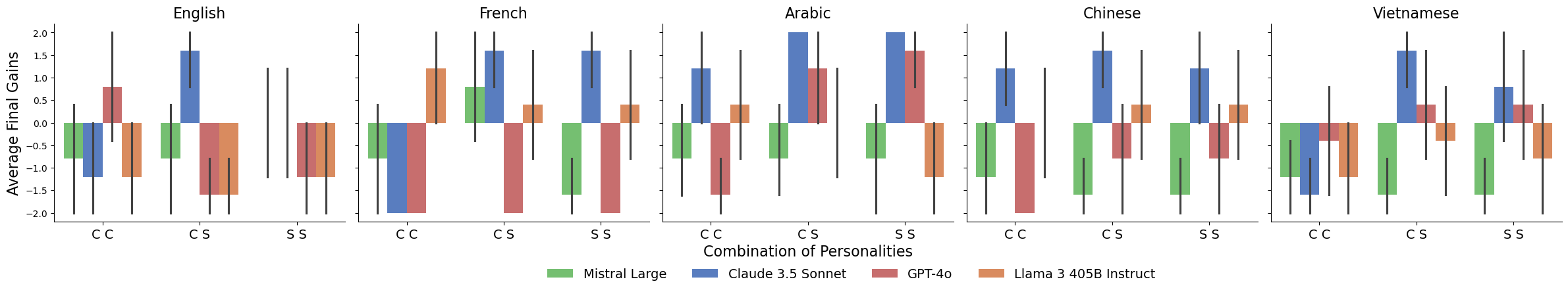}
\end{center}
\caption{Final payoffs of agent 1 in the one-shot zero-sum game, for each LLM (see legend for colour-coding), combination of personalities (columns) and language (rows). }
\label{fig:zero-sum}
\end{figure}

The results for the zero-sum game are reported in Fig. \ref{fig:zero-sum} (we only show the average payoff $P_1$ of agent 1 over the repeated experiments; the payoff for agent 2 its complement to 0, by definition of the game). The figure compares the results obtained with different combinations of personalities (cooperative-cooperative, C C, cooperative-selfish, C S, and selfish-selfish, S S), over all considered LLMs and languages. 

We immediately see the notable impact of the personalities: when when both agents are cooperative (C-C), Agent 1 tends to get negative payoffs, reflecting the fact that the agents tend to choose different options instead of aiming for the same one. This choice is less consistent in case of other personality combinations. Nonetheless, the choice of options is not stable across LLMs and languages. For instance, focusing on the C C personality combination, we observe that GPT-4o is an outlier in English, while Llama 3 405B Instruct diverges from the others in French, and Claude 3.5 Sonnet drastically differs from other LLMs in Arabic and Chinese. Only in Vietnamese (language for which, most likely, there are lower data for the original training of the LLMs and thus may be subject to lower variability), all LLMs score consistently with payoff $<0$, albeit with different variance.

Similar observations hold for the other personality combinations, across languages: overall, there is great variability and hardly recognised conserved patterns, and the LLMs seldom agree with one another, or are even consistent with themselves, when the language is changed. According to literature, the best strategy for a zero-sum game is a mixed strategy (or, in the one-shot case, even a random choice); however, it seems that each LLM chooses sometimes consistently for each combination of language and personality (note that the credible interval bars are very small in some cases, such as C C in French for GPT-4o) and other times in rather random fashion (e.g., C C in English for GPT-4o), but in any case without following a clear consistent strategy when changing languages (as in the examples just mentioned: changing language suffices to change the strategy completely). All in all, these observations should warn about the choice of LLMs to be used for cybersecurity applications, as they may be extremely sensitive about geographical location and language, as well as on other characteristics of the LLM agents that can be defined by the developer of by the user. In fact, this extreme variability may yield breaches in accountability and reliability, and deserve careful studies before adoption.

To go beyond qualitative investigation, we use the metrics defined in Sec. \ref{sec:metrics} to quantitatively compare the LLMs, and help to guide their selection.
Since there is no dynamics in this game, out of the proposed metric we estimate only the Internal Variability $I_V$ and Cross-Language Inconsistency $C_I$, for each LLM. The results are reported in Table \ref{tab:zero_scores}. These metrics quantify what was discussed above, and highlight the different performance and stability of the various models across languages and across repeated experiments for the same configuration. Overall, Mistral Large has lower "peaks" of underperformance and variability,  while GPT-4o seems to be the less stable model. Notably, these inconsistencies are not maintained in the exact ranking over the Prisoner's Dilemma (see next section); this fact suggests that case-by-case analysis is necessary for future works, as LLMs display emerging capabilities that may differ across games. Choosing the best LLM to apply cybersecurity protocols is thus a delicate endeavour that will require dedicated studies and protocols.

\begin{table}[ht]
\centering
\caption{Internal Variability (IV) and Cross-Language Inconsistency (CI) metrics for the zero-sum game across LLMs. Lower values indicate more stable and consistent model behaviour.
}
\vspace{3mm}
\begin{tabular}{c|c|c|c|c}
& Mistral Large & Claude 3.5 Sonnet & GPT-4o & Llama 3 405B Instruct \\
\hline
$I_V$ & 0.87 & 1  & 0.79 & 0.90  \\
$C_I$ & 0.29 & 0.58  & 1 & 0.46 \\
\end{tabular}
\label{tab:zero_scores}
\end{table}

\subsection{Repeated Prisoner's Dilemma}

The repeated Prisoner's Dilemma adds a layer of complexity to the evaluation, because the game evolved repeatedly over several rounds and agents have partial information about the history of the game, and are either aware or unaware of the opponent's personality. As such, they can make conditional decisions on  the accessible  history. The following results can be further complemented by results in \cite{buscemi2025fairgame}, which present a broader outlook onto LLM-based games.

\begin{figure}[h!]
\begin{center}
\includegraphics[width=15cm]{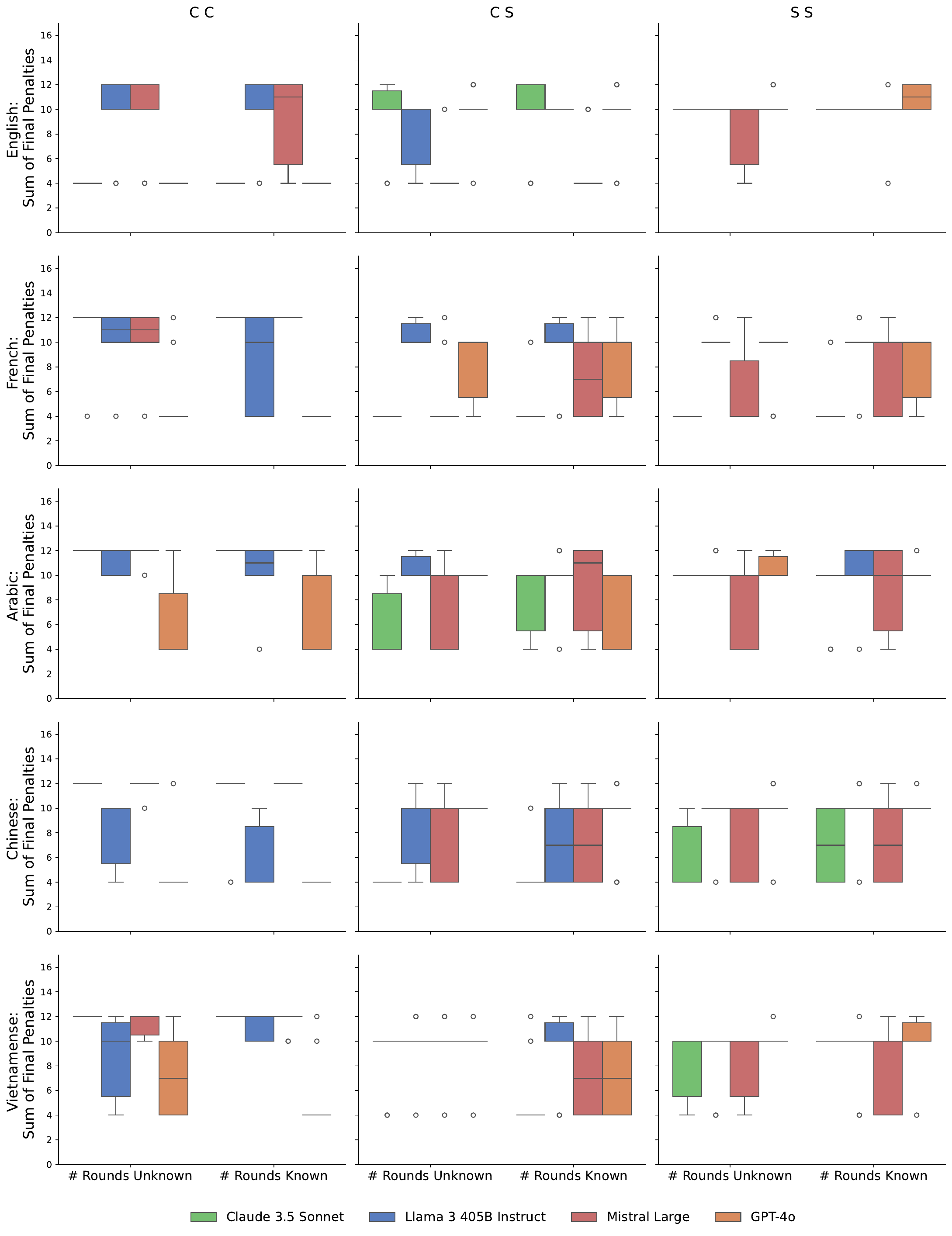}
\end{center}
\caption{Aggregated final payoffs of the repeated Prisoner's Dilemma games over repeated experiments, for each LLM (see legend for colour-coding), combination of personalities (columns), language (rows), and knowledge of opponent's personality (x-axis). }
\label{fig:prisoner_all}
\end{figure}

Fig. \ref{fig:prisoner_all} shows the box plots for the final payoff (representing penalties) for the agents, with quartiles of the payoff distribution. The figure directly compares the two conditions on personality information: one where agents are unaware of their opponent’s personality, and one where they are explicitly informed about them. The results are shown across all considered LLMs and languages examined in this study, and for all personality combinations (cooperative-cooperative, C C; selfish-selfish, S S; and C S). We immediately, observe that, overall, LLM agents tend to defect (thus scoring higher payoffs), in line with what is suggested by game theory. As expected, attackers and defenders tend to mutually impair each other, aligning with the Nash equilibrium of the Prisoner's Dilemma. However, notable exceptions exist, and there are dramatic inconsistencies across languages and combinations of personalities; this indicates that, on top of the payoff matrix, languages and intrinsic biases may influence the agents' behaviour. 

When focusing on the individual features, we see that some LLM are more "stable" than others, that is, they provide similar outputs across languages: Llama 3 and GPT-4o, overall, produce similar distributions in payoffs (even though discrepancies exist when playing the game in one language or another, see  \textit{e.g.,} that GPT-4o C-S players tend to have lower penalties (thus cooperate more) when playing in French than in Arabic or Mandarin Chinese. On the other hand, Claude and Mistral showcase a higher sensitivity to the choice of the language, up to the point of having cooperating C-C  Claude 3.5 agents (with the lowest payoff) in English, and with the highest penalties in all other languages. In general, penalties are lower in English and when the number of rounds is unknown, indicating more consistent cooperative behaviour in the LLM primary training language. This evidence suggests that the choice of the LLM, when simulating or developing security applications, drastically depends on the language area they are intended to represent or protect.

Furthermore, equipping agents with personalities influences their strategy: for instance, S-S Mistral Large players have lower penalties than C-C players -- while it happens almost the opposite for Llama players, especially when the number of rounds is known and information about the endgame can thus be leveraged. Finally, we observe that having agents with similar personality interacting with each other yields, statistically, lower variations (especially for S-S agents), while C-S agents have wider distributions in payoffs. These observations suggest that the higher flexibility bestowed upon agents built with generative AI also leads to emerging and potentially unpredictable behaviours. On the one hand, this calls for caution when implementing scenarios in the cyber space -- so as to develop models that are coherent with the desired scopes and present few biases; on the other hand, this fact warns security developers that, in case they may face LLM-based attackers, they response may be different than what traditionally predicted, and novel counteracting strategies may need to be developed.

\begin{figure}[h!]
\begin{center}
\includegraphics[width=10cm]{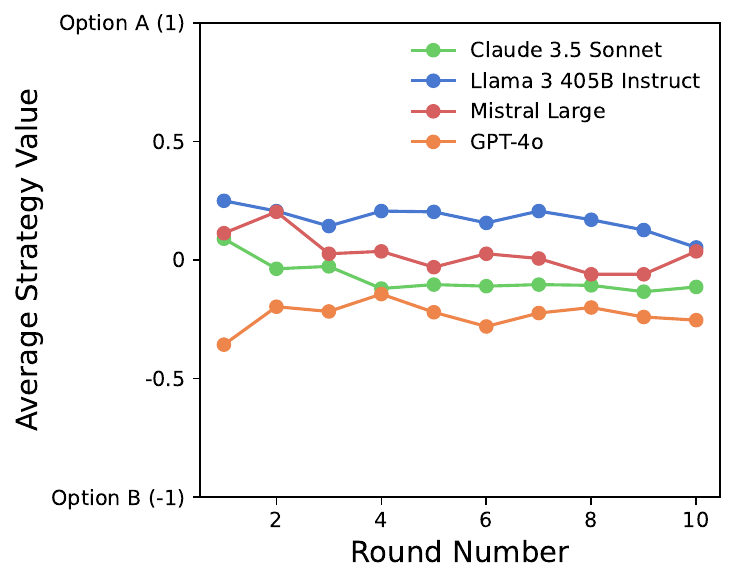}
\end{center}
\caption{Evolution of normalized penalties (averaged over repeated experiments) over repeated rounds, for each LLM within the Prisoner' Dilemma scenario. }
\label{fig:prisoner_runs}
\end{figure}

To look at how games evolve over the rounds, look at Fig. \ref{fig:prisoner_runs}. We recall that, to enable direct comparison between LLMs, the payoff average results were normalised between minimum and maximum. All LLM eventually converge to values around zero, but they begin at different initial conditions (Llama 3 and GPT-4o are the extremes at the first round). Claude 3.5 Sonnet converges rapidly to stable payoff values within a few rounds. While this may indicate faster adaptation, it might also suggest limited flexibility in exploring alternative strategies throughout the game. Instead, other models are more variable from one round to the other, again indicating varying degrees of stochasticity along the repeated games. The general downward trend in penalties over rounds for Claude 3.5, Llama 3.1 405B and Mistral Large indicate progressively increasing mutual cooperation among agents; this is consistent with the strategies traditionally observed in repeated games, where agents reciprocate cooperation to maximize long-term payoffs \cite{wang2015universal}. Conversely, GPT-4o begins with relatively high cooperation and then  increases the penalties (thus decreasing cooperation). This reflects potential biases towards cooperative behaviours in the case of one-shot Prisoner's Dilemma game (at round one), eventually balanced by context-dependent strategic adaptation. With these results, we thus observe that agents perform behaviours on top of what is purely predicted by the payoff matrix, and that repeated interactions yield different results than the one-shot counterparts. 

\begin{figure}[h!]
\begin{center}
\includegraphics[width=10cm]{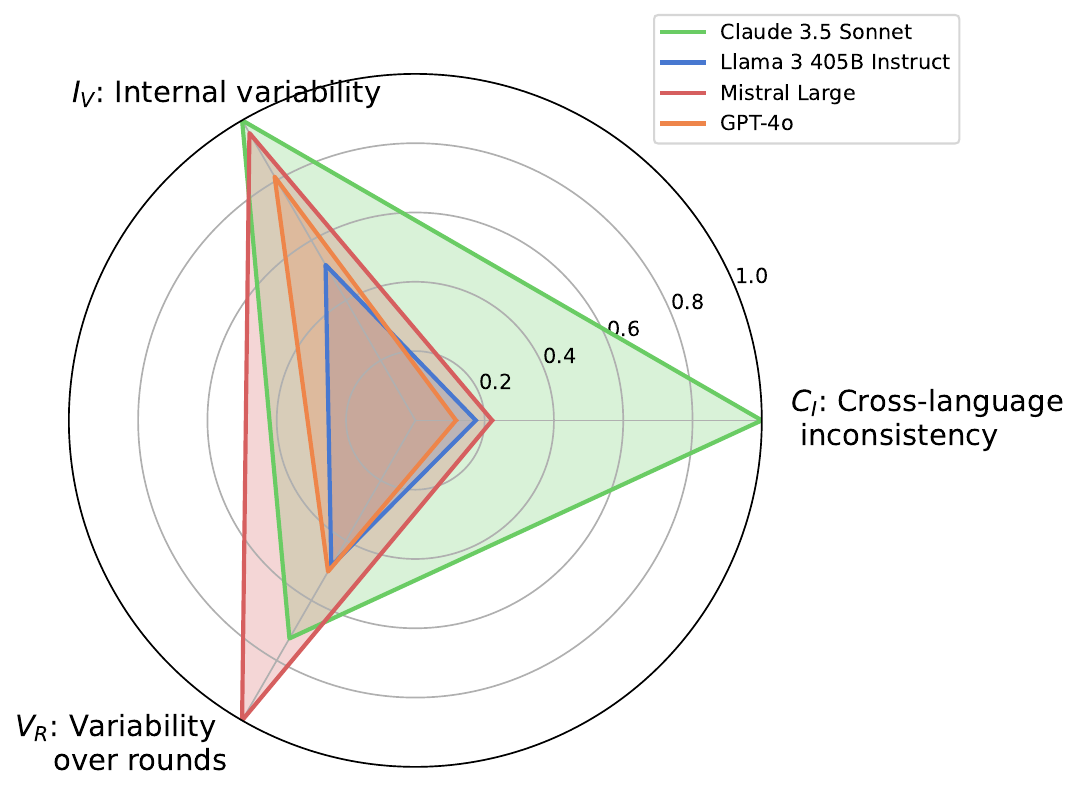}
\end{center}
\caption{Radar plot mapping the three metrics described in Sec. \ref{sec:metrics}, for Prisoner's Dilemma and for all considered LLMs.}
\label{fig:prisoner_spider}
\end{figure}

What has been qualitatively described above is quantitatively captured in Fig. \ref{fig:prisoner_spider}, which sums up the metrics used to measure, for each LLM, the variability across repeated experiments inconsistencies across languages, and variability during repeated games (see Sec. \ref{sec:metrics}). Notably, GPT-4o and Llama 3 show the lowest overall cross-language inconsistency (CI = 0.37 and CI = 0.42 respectively), while Claude 3.5 exhibits the highest CI (0.79), suggesting a higher sensitivity to prompting language. Moreover, we immediately recognise the higher variability displayed by Claude 3.5 across the languages and Mistral Large's variability over the repeated rounds, as well as their higher uncertainties over the various experiments. Conversely, GPT-4o and Llama 3 show more consistent results, indicating some stabilising effect that somehow copes with their stochastic behaviour.

\section{Discussion}

Real-world cyber systems are characterised by higher complexity (\textit{e.g.,} partial information or resources, adaptive infiltration schemes, uncertainties) that may divert agents to always perform best-payoff actions. Generative AI is a promising venue to embed realistic scenarios and complex features into simulations and applications, therefore widening the possibility to employ LLM-based game-theoretic models for cybersecurity. However, as LLMs are emerging technologies with unpredictable and often un-interpretable capabilities, it is imperative to systematically assess their capabilities and behaviours. This study provides evidence that LLM agents may behave sub-optimally in key games used for cybersecurity applications, highlighting that the language used for prompting the models, as well as additional traits such as completeness of information or the assigned digital personality of agents may introduce behavioural biases that affect their decision-making during the games. 

Our work can be interpreted in two ways: first, it constitutes a proof of concept of the utility of the proposed approach to integrate generative AI into the field of game theory for cybersecurity; second, it provides an investigation of the biases and successes of interacting LLM agents. Despite being limited to two classes of $2\times 2$ games, based on simplified assumptions that allowed the comparison of outcomes stemming from various bias sources, our study already recognises several sources of ambiguity in LLM responses, paving the way to future studies focused on specific applications and mitigation of LLM issues. Future works may also test additional games, such as Stackelberg games, Markovian games, or evolutionary games, and increase the degrees of freedom associated with playing agents, \textit{e.g.,} by equipping them with complex personalities or different degrees of information, as well as consider multi-agent games on networks. Building upon our work, broad investigations can thus be conducted.

While our study offers meaningful insights into LLM-driven game-theoretic behaviour, it is not without limitations. To begin with, we focused on only two classes of 2×2 games, namely the zero-sum game and the repeated Prisoner’s Dilemma; although canonical games, they do not fully capture the complexity and breadth of real-world cybersecurity scenarios. Additionally, the selection of five languages, while covering several major linguistic families, does not exhaust the full spectrum of cultural and linguistic variation. Lastly, the experiments were conducted in simulation without real-world network deployments or adversarial environments, leaving open the question of how these models would perform in operational cybersecurity settings. These relevant questions may constitute basis for future work.

Overall, we observed that, despite the great promises of generative AI to positively impact the development of security applications in the cyber domain (as outlined, \textit{e.g.,} by \cite{he2025generative} when implementing robust mobile networking), LLMs still face notable limitations in handling uncertainty, strategic planning capabilities, and sensitivity to embedded biases. Our methodology and case studies suggest that, before being routinely applied, generative algorithms should be carefully tested by the community in a variety of scenarios and by considering numerous features. Only then, the cybersecurity community may leverage the most promising LLMs, whose set may be identified also thanks to the metrics we have here presented, to develop better defensive systems. 

Indeed, the use of LLMs in cybersecurity contexts raises important ethical considerations. Simulating attacker and defender behaviours with AI-driven agents may enable better preparation and defence mechanisms, but it also opens the door to malicious uses, such as automated vulnerability discovery or adversarial prompt engineering. Moreover, biases in LLM behaviour, especially when influenced by language or personality traits, could lead to unintended consequences in sensitive security applications. As such, we advocate for responsible experimentation frameworks and transparency in reporting LLM-driven cybersecurity simulations.
In fact, our case studies point to potential vulnerabilities that need to be carefully considered: if used maliciously, LLMs may behave differently from other traditional algorithms (for instance, by altering cooperative behaviours depending on the language) and bypass solutions tested on more traditional scenarios. This observation thus calls for renewed attention towards these emerging technologies, and suggests the use of coherent testing frameworks, such as FAIRGAME, to systematically test scenarios of increasing complexity. Overall, such tests would enrich our understanding of LLM behaviours in the cyber systems and would help make better predictions and interventions to navigate the newest technologies.

\section*{Conflict of Interest Statement}

The authors declare that the research was conducted in the absence of any commercial or financial relationships that could be construed as a potential conflict of interest.

\section*{Author Contributions}

\textbf{D.P.}: conceptualization, formal analysis, investigation, methodology, validation, visualization, writing—original draft, writing—review and editing. \textbf{A.B.}: investigation, software, methodology, visualization, writing—original draft, writing—review and editing. \textbf{A.D.S.}: methodology, validation, writing—review and editing. \textbf{T.A.H.:} methodology, validation, writing—review and editing. \textbf{G.C.:} funding acquisition, supervision, writing—review and editing. \textbf{P.L.:} methodology, supervision, writing—review and editing.

\section*{Funding}
Details of all funding sources should be provided, including grant numbers if applicable. Please ensure to add all necessary funding information, as after publication this is no longer possible.

\section*{Acknowledgments}
The authors would like to thank the native speaker colleagues who helped translating the scenarios.


\section*{Data Availability Statement}
The code for this study can be found on Github at \texttt{https://github.com/aira-list/FAIRGAME}, together with examples of configuration files and templates.

\bibliography{arxiv_version}


\newpage
\section*{Figure captions}


\end{document}